\documentclass[11pt]{article}
\usepackage{graphicx}

\newsavebox{\hflrar}
\sbox{\hflrar}{\makebox[0pt][l]
{${\scriptstyle \leftharpoonup}$}{${\scriptstyle \rightharpoonup}$}}

\def \to {\rightarrow}

\begin{document}
\begin{center}
{\Large\bf Transverse-Momentum Dependent Factorization for $\gamma^*
\pi^0 \to \gamma $} \vskip 10mm
J.P. Ma$^1$ and Q. Wang$^{1,2}$    \\
{\small {\it $^1$ Institute of Theoretical Physics, Academia Sinica,
Beijing 100080, China }} \\
{\small {\it $^2$ Department of Physics and Institute of Theoretical
Physics, Nanjing Normal University, Nanjing, Jiangsu 210097, P.R.China}} \\
\end{center}

\vskip 1cm
\begin{abstract}
With a consistent definition of transverse-momentum dependent (TMD)
light-cone wave function, we show that the amplitude for the process
$\gamma^* \pi^0 \to\gamma$ can be factorized when the virtuality of the
initial photon is large. In contrast to the collinear factorization
in which the amplitude is factorized as a convolution of the
standard light-cone wave function and a hard part, the TMD
factorization yields a convolution of a TMD light-cone wave
function, a soft factor and a hard part. We explicitly show that the
TMD factorization holds at one loop level. It is expected that the
factorization holds beyond one-loop level because the cancelation of
soft divergences is on a diagram-by-diagram basis. We also show that
the TMD factorization helps to resum large logarithms of type
$\ln^2x$.
\vskip 5mm \noindent
\end{abstract}
\vskip 1cm
\par\vfil
\eject

\noindent
{\bf 1. Introduction}
\par
Factorization is an important tool to make predictions from QCD for
an exclusive process if it contains both long- and short-distance
effects, the latter involving one or more large momentum transfers,
denoted generically as $Q$. It has been proposed long time ago that
the amplitude of such a process can be factorized by an expansion of
the amplitude in $1/Q$, corresponding to an expansion of QCD
operators in twist\cite{BL,CZrep}. The leading term can be
factorized as a convolution of a hard part and light-cone wave
functions of hadrons. The light-cone wave functions are defined with
QCD operators, and the hard part describes hard scattering of
partons at short distances. In this factorization the transverse
momenta of partons in parent hadrons are also expanded in the hard
scattering part. At leading twist they are neglected. This results
in that the light-cone wave functions depend only on the
longitudinal momentum fractions carried by partons, not on
transverse momenta of partons. This is the so-called collinear
factorization.
\par
The collinear factorization has been also used for inclusive
processes. At leading twist the transverse momenta of partons are
also neglected, and the resulting standard parton distributions only
depend on longitudinal momentum fractions. It is interesting to note
that a transverse-momentum dependent(TMD) factorization can be made
by introducing TMD parton distributions \cite{CS,CSS,JMY}. With a
consistent definition, TMD parton distributions have some
interesting relations to the standard Feynman distributions. For
example, non-light-like gauge links shall be used to avoid some
inconsistencies as discussed in the detail in \cite{Col1}. The TMD
factorization has many advantages. Beside those for different
aspects of relevant physics, a prominent advantage is that it leads
to the so-called CSS resummation formalism for resumming large
logarithms appearing in the collinear
factorization\cite{CS,CSS,JMY}. This implies that the two
factorization approaches are closely related to each other.
\par
In the collinear factorization for an exclusive process one usually
has large contributions from end-point regions where the
longitudinal momentum fraction $x$ is small or close to $1$,
signified by large logarithms $\ln^2x$ in the perturbative expansion
of the hard part. As such the expansion converges slowly and
predictions can not be reliably made by taking few leading terms in
the expansion only. It has been shown that those large logarithms
can be resummed by using TMD- or $k_T$- factorization\cite{BS,LS}.
Similarly to inclusive cases, one introduces TMD light-cone wave
functions (LCWF) by using non-light-like gauge links. Recently, such
a factorization has attracted many attentions related to studies of
exclusive $B$-meson decays. For these decays, one can use collinear
factorization to factorize short- and long-distance effects
\cite{BBNS}. Beside large logarithms, there are even end-point
singularities in the factorization of certain decays. This makes the
factorization at the end-point useless. In \cite{KTB} it has been
suggested to use $k_T$- or TMD factorization to resum large
logarithms or to eliminate those end-point singularities.
\par
Comparing to TMD factorization for inclusive processes which has
been studied extensively (see \cite{CS,CSS,JMY} and references
therein), the TMD or $k_T$-factorization for exclusive processes
have been not studied with consistent definitions of TMD LCWF's as
much as that for inclusive processes. For a given process the
amplitude in collinear factorization is factorized in terms of a
hard part and LCWF's, latter representing all nonperturbative
effects in the process. In \cite{LS,KTB} the TMD factorization can
be taken as a "direct" generalization of the results of collinear
factorization in the sense that the amplitude is factorized as a
convolution only with a hard part and TMD LCWF's. From the study of
TMD factorization for inclusive processes\cite{CS,CSS,JMY} one knows
that the differential cross-section for a given process can be
factorized as a convolution of a hard part, TMD parton distributions
and {\it a soft factor}. The nonperturbative effects in the
inclusive process are not only represented by the TMD parton
distributions, but also by the soft factor. This is in contrast to
the collinear factorization, where the differential cross-section is
a convolution of a hard part and standard parton distributions.
Therefore, TMD factorization with just TMD LCWF's as nonperturbative
objects in an exclusive process may be incomplete and some extra
nonperturbative objects like soft factors may be needed to complete
the factorization. It should be noted that the possible existence of
soft factors in the factorization can change the scale evolutions
which are used to resum large logarithms. One needs to examine the
factorization in detail.
\par
In the study of TMD factorization for radiative leptonic decay of
$B$-meson\cite{MW2}, it has been shown the decay amplitude takes the
convolution form, with a hard part, the TMD LCWF of $B$-meson, and a
soft factor. The TMD LCWF of $B$-meson has been consistently defined
and its relation to the standard LCWF has been studied in detail in
\cite{MW1}. From the result in \cite{MW2} the soft factor is really
needed in order to make the factorization complete. A complete
factorization means that the hard part does not contain any soft
divergence like collinear- and infrared (I.R.) divergence. It has
been shown that two factorizations for this simple case are
equivalent and large logarithms can be conveniently resummed.
\par
In this work we study TMD factorization for the process $\gamma^*
\pi \to \gamma$, where only one light hadron is involved. With the
present study and that in \cite{MW2,MW1}, we have examined TMD
factorization for an exclusive process involving one hadron. This
can be thought as a first step to the goal to examine TMD
factorization for processes involving more than one light hadrons
and to study the factorization for $B$-decays into light hadrons. A
consistent definition of the TMD LCWF of $\pi$ was first given in
\cite{BS}. The TMD LCWF has been studied in\cite{JMW}, where some
interesting relations to the standard LCWF have been obtained.
We will consider the case that the virtual photon has the negatively large virtuality
$-Q^2 (Q^2>0)$, which converts $\pi$ into a real photon.
Because of the large $Q^2$, one can expand the amplitude in the inverse
of $Q^2$.
We will show that the amplitude of $\gamma^* \pi \to \gamma$ at the leading order
of the inverse $Q^2$ can be
factorized at one-loop level as the convolution
\begin{equation}
  \phi_+ \otimes \tilde S \otimes H  \cdot \left \{ 1 + {\mathcal O }(\Lambda^2 /Q^2) \right \}.
\end{equation}
In the above $\phi_+$ is the TMD LCWF, $\tilde S$ is the soft factor
and $H$ is a coefficient function which can be safely calculated in
perturbative QCD. The soft factor is defined with gauge links and it
is needed to ensure $H$ free from I.R. singularities.
The correction to the factorized form is power-suppressed and is proportional
to $\Lambda^2/Q^2$.
$\Lambda$ is any possible soft scale characterizing nonperturbative
scale, like $\Lambda_{QCD}$, the mass of $\pi$ etc.. This scale
is about several hundreds MeV. Therefore,the correction to the
factorized form is small with the large $Q^2$.
\par
In the
factorized form, the TMD light-cone wave function only depends
on the specific hadron and is universal for all possible processes.
The coefficient function $H$ depends on the process but does not
depend on the specific hadron. It represents a hard transition
of a $q\bar q$ with the virtual photon into the real photon,
in which the pairs move along one light-cone direction and
has a small relative  transverse momentum.
The coefficient function $H$ is obtained from the transition
amplitude where the nonperturbative effects represented
by soft divergences are subtracted.
The soft-factor depends neither on the process
nor on the specific hadron, it is completely determined
by the infrared behavior of QCD.
The TMD light-cone wave function
and the soft factor are nonperturbative objects and can only be studied
with nonperturbative methods of QCD, like Lattice QCD or QCD sum rule method.
Phenomenologically, they can also be extracted
from relevant processes. In the case of a pion, they can be extracted, e.g.,
from the process studied here, and those like $\pi$ form factor,
exclusive decays of $B$-mesons, etc... However, beside the process studied here,
one needs first to show that the TMD factorization for other processes
can be consistently
performed.
In this
work we will show that the proposed factorization for $\gamma^* \pi \to \gamma$
holds at one-loop
level and the cancelation of all soft divergences is on a
diagram-by-diagram basis. The later is important to extend the
factorization beyond one-loop level. We will also show that the
factorization provides a simple way to resum large logarithms.
\par
Our paper is organized as the following: In Sect.2 we give the
definition of TMD LCWF and its one-loop results which will be used
to show the factorization. In Sect.3 we introduce our notation and
show the TMD factorization at tree level. In Sect.4 we show that the
TMD factorization can be completed at one-loop level by introducing
a soft factor. The soft factor is defined as products of gauge
links. In Sect.5 we show that possible large logarithms around
end-point regions can be resummed with our method. Sect.6 contains a
summary.
\par\vskip20pt
\noindent
{\bf 2. The Definition of TMD LCWF and Its One-loop Results}
\par
We will use the  light-cone coordinate system, in which a
vector $a^\mu$ is expressed as $a^\mu = (a^+, a^-, \vec a_\perp) =
((a^0+a^3)/\sqrt{2}, (a^0-a^3)/\sqrt{2}, a^1, a^2)$ and $a_\perp^2
=(a^1)^2+(a^2)^2$. We
introduce a vector $u^\mu=(u^+,u^-,0,0)$ and the definition of
TMD LCWF for $\pi^0$ can be given as
in the limit $u^+ << u^-$:
\begin{eqnarray}
\phi_+(x, k_\perp,\zeta, \mu) &=& \ \int \frac{ d z^- }{2\pi}
  \frac {d^2 z_\perp}{(2\pi )^2}  e^{ik^+z^- - i \vec z_\perp\cdot \vec
k_\perp}
\langle 0 \vert \bar q(0) L_u^\dagger (\infty, 0)
  \gamma^+ \gamma_5 L_u (\infty,z) q(z) \vert \pi^0(P) \rangle\vert_{z^+=0},
\nonumber\\
   k^+ &=& xP^+, \ \ \ \ \  \zeta^2 = \frac{2 u^- (P^+)^2}{u^+}\approx
\frac{ 4 (u\cdot P)^2}{u^2},
\label{def}
\end{eqnarray}
where $q(x) $ is the light-quark field.
$L_u$ is the gauge link in the direction $u$:
\begin{equation}
L_u (\infty, z) = P \exp \left ( -i g_s \int_{0} ^{\infty} d\lambda
     u\cdot G (\lambda u+ z ) \right ) .
\end{equation}
The definition has been given in \cite{BS}. In the above, $\pi^0$
has the momentum $P^\mu =(P^+, P^-,0,0)$. The limit $u^+<< u^-$
should be understood that we do not take the contributions
proportional to any positive power of $u^+/u^-$ into account. The
definition is gauge invariant in any non-singular gauge in which the
gauge field is zero at space-time infinity. It has no light-cone
singularities as we will show through our one-loop result, but it
has an extra variable $\zeta^2$. The evolution with the
renormalization scale $\mu$ is simple:
\begin{equation}
\mu \frac{\partial \phi_+(x, k_\perp,\zeta, \mu) }{\partial \mu }
= 2\gamma_q  \phi_+(x, k_\perp,\zeta, \mu),
\end{equation}
where $\gamma_q$  is the anomalous dimension of the light quark field
$q$ in the axial gauge $u\cdot G=0$.
In the above we do not specify if the light quark $q$ is a $u$- or
$d$-quark.
The definition
can be generalized to other light hadrons and there can be various
TMD LCWF's. These TMD LCWF's of various hadrons have been classified
in \cite{JMYWF}, and their asymptotic behavior with $k_\perp\to\infty$
has been derived\cite{JMYPC}.
\par
To perform TMD factorization one needs to calculate the wave function
with perturbative QCD, in which $\pi^0$ is replaced by a partonic
state. We take the partonic state $\vert q(k_q), \bar
q(k_{\bar q})\rangle$ to replace $\pi^0$ in the definition,
the momenta are given as
\begin{equation}
k_q^\mu =(k_q^+, k_q^-,\vec k_{q\perp}), \ \ \ \  k_{\bar q}^\mu
=(k_{\bar q}^+, k_{\bar q}^-, -\vec k_{\bar q\perp}), \ \ \ \ \ \
k_q^+ = x_0 P^+, \ \ \ \  k_{\bar q}^+ = (1-x_0) P^+ = \bar x_0 P^+.
\end{equation}
These partons are on-shell. In this work we will keep the quark mass
$m$ for light quarks to regularize collinear singularities and give
a small gluon mass $\lambda$ to regularize I.R. singularities. We
will omit the $\mu$-dependence in various quantities in this work
and this dependence is always implied.
\par
At tree-level, the wave function reads:
\begin{equation}
\phi_+^{(0)} (x,k_\perp,\zeta) =  \delta (x-x_0)
     \delta^2(\vec k_\perp -\vec k_{q\perp}) \phi_0, \ \ \ \ \  \phi_0 =\bar
v(k_{\bar q}) \gamma^+ \gamma_5 u(k_q) /P^+.
\end{equation}
We will always write a quantity $A$ as $A=A^{(0)} + A^{(1)} +\cdots$, where
$A^{(0)}$ and
$A^{(1)}$ stand for tree-level- and one-loop contribution respectively.
At one-loop one can divide the corrections into a real part and a virtual
part.
The real part comes from contributions of Feynman diagrams given in Fig.1.
The virtual part comes from contributions of Feynman diagrams given in
Fig.2.,
those contributions are proportional to the tree-level result.
\par
\begin{figure}[hbt]
\begin{center}
\includegraphics[width=11cm]{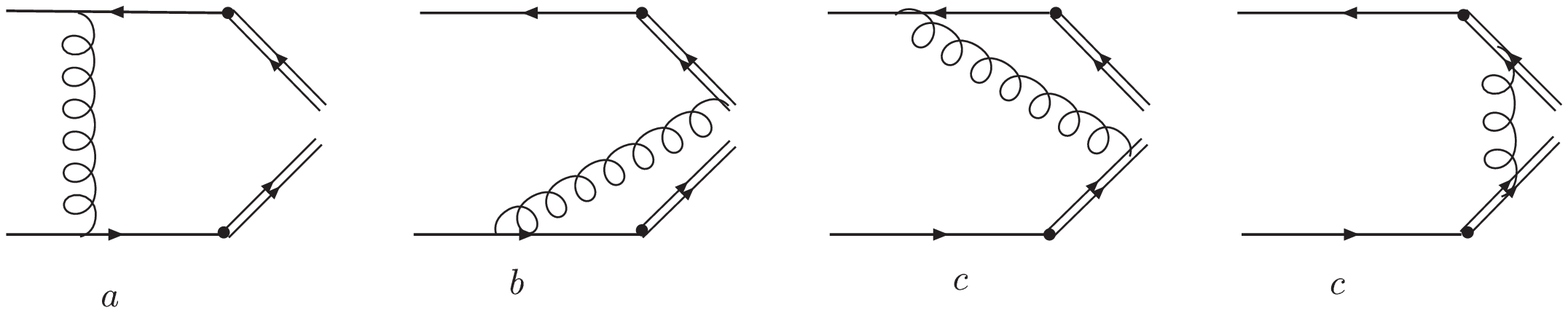}
\end{center}
\caption{The real part of one-loop contribution to the TMD LCWF.
The double lines represent the gauge links.  }
\label{Feynman-dg1}
\end{figure}
\par
The contributions from Fig.1b, Fig.1c and Fig.1d read:
\begin{eqnarray}
\phi_+(x,k_\perp,\zeta)\vert_{1b} &=& -\frac{ \alpha_s C_F}{2\pi^2}
\left \{ \theta (x_0 -x) \left ( \frac{1}{x-x_0}\right)_+ \frac{x}{x_0
\Delta_q} +
\frac{1}{2} \delta (x-x_0) \frac{1}{q_\perp^2 +\lambda^2} \ln\frac{q_\perp^2
+\lambda^2}{\zeta^2 x_0^2}
\right \} \phi_0,
\nonumber\\
\phi_+(x,k_\perp,\zeta)\vert_{1c} &=& \frac{ \alpha_s C_F}{2\pi^2}
\left \{ \theta (x -x_0) \left ( \frac{1}{x-x_0}\right)_+ \frac{1-x}{(1-x_0)
\Delta_{\bar q}}
- \frac{1}{2} \delta (x-x_0) \frac{1}{q_\perp^2 +\lambda^2}
\ln\frac{q_\perp^2 +\lambda^2}{\zeta^2 \bar x_0^2}
\right \} \phi_0,
\nonumber\\
\phi_+(x,k_\perp,\zeta)\vert_{1d} &=& - \frac{ \alpha_s C_F}{2\pi^2}
\frac{1}{\lambda^2 + q^2_\perp}\delta(x-x_0) \phi_0,
\end{eqnarray}
with
\begin{eqnarray}
\Delta_q &=& (1-y)^2 m^2 + y\lambda^2 + (k_\perp - y k_{q\perp})^2, \ \ \
y= \frac{x}{x_0}, \ \ \ \  q_\perp = k_\perp - k_{q\perp}.
\nonumber\\
\Delta_{\bar q}  &=& \left ( 1 -\bar y   \right )^2 m^2 + \bar y \lambda^2
+ \left ( k_\perp  - \bar y  k_{q\perp} \right )^2, \ \ \
\bar y = \frac{\bar x}{\bar x_0}, \ \ \ \  q_\perp = k_\perp - k_{q\perp}.
\label{1LWF}
\end{eqnarray}
It should be noted that if we take $u^+=0$ in the definition at the
beginning there are light-cone singularities, represented by terms
like $1/x$ and $1/(1-x)$. With a nonzero, but small $u^+$ these
singularities are regularized. All these contributions behave like
$k_\perp^{-2}$ when $k_\perp$ becomes large. The contribution from
Fig.1a is too lengthy. But its calculation is standard because the
gauge link is not involved. The contribution also behaves like
$k_\perp^{-2}$ when $k_\perp$ becomes large. The leading
contribution for large $k_\perp$ is:
\begin{eqnarray}
\phi_+(x,k_\perp, \zeta)\vert_{1a}  &=& \frac{2\alpha_s}{3\pi^2}
\frac{1}{ k_\perp^2 +\Lambda^2 } \left (\frac{x}{x_0} \theta (x_0-x)
    + \frac{1-x}{1-x_0} \theta (x-x_0)\right ) \phi_0
\nonumber\\
   && + {\rm ``finite\ terms "},
\label{1LWFA}
\end{eqnarray}
where we present only the U.V. divergent part explicitly and
introduce a ``cut-off" $\Lambda$, which also appear in the finite
term so that the sum does not depend on it. All terms in the second
line are proportional $k^{-4}_\perp$ when $k_\perp \to \infty$.
Therefore, the one-loop correction of $\phi_+(x,k_\perp, \zeta)$ is
proportional to $k^{-2}_\perp$ when $k_\perp \to \infty$, as
expected from the general power counting rule \cite{JMYPC}.
\par
\begin{figure}[hbt]
\begin{center}
\includegraphics[width=8cm]{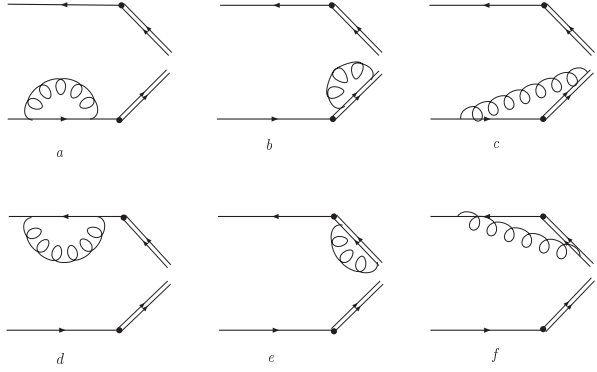}
\end{center}
\caption{The virtual part of one-loop contribution to the TMD LCWF.
The double lines represent the gauge links.  }
\label{Feynman-dg2}
\end{figure}

\par
The virtual part of the one-loop correction is from the Feynman
diagrams in Fig.2. Contributions from each diagrams are:
\begin{eqnarray}
\phi_+ (x,k_\perp, \zeta)\vert_{2a} &=&\phi_+ (x,k_\perp, \zeta)\vert_{2d}
= \phi_+^{(0)} (x,k_\perp, \zeta) \cdot \frac{\alpha_s}{6\pi}
    \left [ -\ln \frac{\mu^2}{m ^2} +2 \ln\frac{m ^2}{\lambda^2} -4 \right ]
,
\nonumber\\
\phi_+ (x,k_\perp, \zeta)\vert_{2b} &=&
    \phi_+ (x,k_\perp, \zeta)\vert_{2e} = \phi_+^{(0)} (x,k_\perp, \zeta)
     \cdot \frac{\alpha_s}{3\pi}
    \ln \frac{\mu^2}{\lambda^2},
\nonumber\\
\phi_+ (x,k_\perp, \zeta)\vert_{2c} &=&
\phi_+^{(0)} (x,k_\perp, \zeta)
\cdot \frac{\alpha_s}{6\pi}
    \left [ 2\ln\frac{\mu^2}{m^2} + 2\ln\frac{\zeta^2 x_0^2}{m^2}
       +\ln^2 \frac{m^2}{\lambda^2} -\ln^2 \frac{\zeta^2 x_0^2}{\lambda^2}
     -\frac{2\pi^2}{3} +4 \right ],
\nonumber\\
\phi_+ (x,k_\perp, \zeta)\vert_{2f} &=& \phi_+^{(0)} (x,k_\perp, \zeta)
\cdot \frac{\alpha_s}{6\pi}
    \left [ 2\ln\frac{\mu^2}{m^2} + 2\ln\frac{\zeta^2 \bar x_0^2}{m^2}
       +\ln^2 \frac{m^2}{\lambda^2} -\ln^2 \frac{\zeta^2 \bar
x_0^2}{\lambda^2}
     -\frac{2\pi^2}{3} +4 \right ].
\end{eqnarray}
The complete one-loop contribution $\phi_+^{(1)}$ is the sum of
contributions from the 10 Feynman
diagrams in Fig.1. and Fig.2.
\par
From the above results one can derive some interesting relations
between the TMD LCWF and standard LCWF and the evolution with
$\zeta$. The evolution with $\zeta$ has been shown to be very useful
for resumming large logarithms. Details about these can be found in
\cite{JMW}.
\par\vskip20pt
\noindent
{\bf 3. TMD Factorization for $\gamma^*\pi^0 \to \gamma$ at Tree-Level}
\par
We consider the process:
\begin{equation}
\pi^0 +\gamma^*\to  \gamma
\end{equation}
where $\pi^0$ carries the momentum $P$ and the real photon momentum
$p$. We take a coordinate system in which these momenta are:
\begin{equation}
P^\mu =(P^+, P^-,0,0), \ \ \ \ \ \  p^\mu =(0,p^-,0,0).
\end{equation}
We will consider the case that the virtual photon has the large
negative virtuality $q^2 =(P-p)^2=-2P^+ p^-=-Q^2$. The process can
be described by matrix element $\langle \gamma (p, \epsilon^*)\vert
\bar q\gamma^\mu q \vert \pi^0(P)\rangle$, which can be
parameterized with the form factor $F(Q^2)$:
\begin{equation}
\langle \gamma (p, \epsilon^*)\vert \bar q\gamma^\mu q \vert \pi^0(P)\rangle
= i \varepsilon^{\mu\nu\rho\sigma} \epsilon^*_{\nu} P_{\rho} p_{\sigma}
F(Q^2).
\end{equation}
For simplicity, we only consider one flavor quark $q$ with electric
charge $1$.
\par
The form factor has been studied in experiments\cite{EXP}.
Theoretically it has been studied in the frame work of collinear
factorization extensively\cite{COF1,COF2,KR,GHM,MR,C1L,C2L}. In
\cite{C1L} the coefficient function in the collinear factorization
has been calculated at one-loop, its two-loop result can be found in
\cite{C2L}. In \cite{KR,GHM,MR} an attempt has been made to take the
transverse momentum into account in order to make predictions more
reliable. In \cite{MR}, following \cite{BS,LS} the resummation of
large logarithms has been performed. However, in these studies, the
issue whether the TMD factorization holds or not has not been
investigated. The main goal of the paper is to address the issue,
and we will not perform a phenomenological study with the
factorization to make any numerical predictions.
\par
To find a factorized form for the form factor, we replace the
hadronic state with the partonic state as before in calculation of
the matrix element. The momenta of the partonic state
is specified in Eq.(5). In general one should
neglect the transverse momentum ${\bf k_{q\perp}}$ in the numerators
in calculating $F(Q^2)$, while
one keeps the most important terms of ${\bf k_{q\perp}}$
in the denominators\cite{LS}. Therefore, what neglected here is proportional
to $k^2_{q\perp}/Q^2$.
However, at tree-level,
the amplitude does not depend on ${\bf k_{q\perp}}$.
At tree-level there are two Feynman diagrams
shown in Fig.3.
\par

\begin{figure}[hbt]
\begin{center}
\includegraphics[width=7cm]{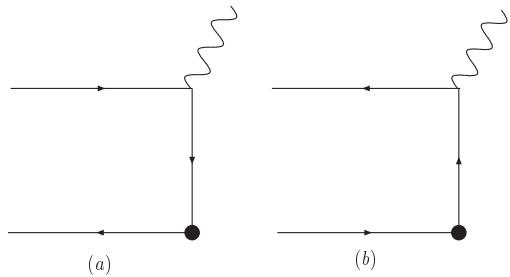}
\end{center}
\caption{Feynman diagrams of tree-level contributions to the
partonic scattering. The black dot denotes the insertion of the
electric current operator corresponding to the virtual photon.}
\label{Feynman-dg5}
\end{figure}
\par
It is straightforward to obtain the tree-level result from Fig.3:
\begin{equation}
F(Q^2)\vert_{3a} = \phi_0 \frac{1}{ Q^2 x_0}, \ \ \ \ F(Q^2)\vert_{3b} =
\phi_0 \frac{1}{Q^2(1-x_0)}=\phi_0 \frac{1}{Q^2\bar x_0}.
\end{equation}
Combining it with the tree-level result of the TMD LCWF, we can
obtain a factorized form for the form factor:
\begin{equation}
F(Q^2) = \frac{1}{Q^2} \int dx d^2 k_\perp \phi_+(x,k_\perp, \zeta) \left [
\frac{1}{x} +\frac{1}{\bar x} \right ].
\end{equation}
\par
As discussed before, such a factorization may not hold beyond
tree-level, and a possible soft factor should be implemented.  We
expect that the form factor in TMD factorization can be written as:
\begin{eqnarray}
F(Q^2) &=& \frac{1}{Q^2 }  \int dx dy  d^2 k_\perp d ^2 l_\perp
\phi_+(x,k_\perp)
          \tilde S (y, l_\perp, \zeta) \theta (x+y ) H (x+y)
          +{\mathcal O}(Q^{-4})
\nonumber\\
      &=& \frac{1}{Q^2} \phi_+ \otimes \tilde S \otimes H + +{\mathcal
O}(Q^{-4}),
\end{eqnarray}
where we only give the dependence on convolution variables explicitly. The
complete
dependence of different objects on different variables will be given in the
next section.
With the above factorization one can easily find the tree-level results
of the hard part $H$ and the soft factor:
\begin{equation}
H^{(0)}(x) = \frac{1}{x} + \frac{1}{1-x}=\frac{1}{x\bar x}, \ \ \ \ \
\tilde S^{(0)}(y, l_\perp ) =\delta(y) \delta^2(\vec l_\perp).
\end{equation}
\par
It should be noted that at tree-level the hard part does not depend
on transverse momenta. This is because we do not have an actual
hadron in the final state of this simple case. When both initial-
and final states contain hadrons, the hard part will depend on
transverse momenta of partons entering hard scattering.
\par\vskip20pt
\noindent
{\bf 4. TMD Factorization at One-Loop Level}
\par
In this section we examine the TMD factorization at one-loop level
and show that the soft factor is indeed needed. We will give its
definition in terms of gauge links. At one-loop level, the
factorization formula in the previous section can be written as:
\begin{equation}
Q^2 F^{(1)} (Q^2) = \phi_+^{(1)}\otimes \tilde S^{(0)}\otimes H^{(0)}
                  +\phi_+^{(0)}\otimes \tilde S^{(1)}\otimes H^{(0)}
                  +\phi_+^{(0)}\otimes \tilde S^{(0)}\otimes H^{(1)}.
\label{con}
\end{equation}
If the factorization is right, $H^{(1)}$ will not contain any soft
divergence and can be then calculated in perturbative theory. If all
soft divergences in $Q^2 F^{(1)} (Q^2)$ are already contained in the
first term, i.e., in $\phi_+^{(1)}\otimes \tilde S^{(0)}\otimes
H^{(0)}$, then there is no need to introduce the soft factor.
However, this is not the case. In this section we will study
different factors introduced above.
\par
In deriving the factorization formula at tree-level, one should
neglect the transverse momentum ${\bf k_{q\perp}}$ in the numerators
in calculating $F(Q^2)$, while
one keeps the most important terms of ${\bf k_{q\perp}}$
in the denominators\cite{LS}.
As discussed in the last section, the hard part at tree level does not
depend on the transverse momenta of partons. In calculating the one-loop
contribution of $F(Q^2)$ with the partonic state,  we will therefore take
the transverse momenta as small quantities and expand it in $k^2_{q\perp}/Q^2$.
We only take the leading order to extract the hard part. The higher orders
are as power-suppressed contributions and they are neglected.

\par\vskip15pt
\noindent
{\bf 4.1. The One-Loop Correction of $F(Q^2)$}
\par
There are 12 Feynman diagrams for the one-loop correction of the
form factor, 6 of them are given in Fig.4. The other 6 diagrams are
obtained from those in Fig.4 by reversing the direction of the quark
line. The diagrams in Fig.4 represent the correction to Fig.3a, and
the other 6 diagrams for the correction to Fig.3b. Two corrections
are related each other by charge conjugation. Hence we will need to
study how the contributions from Fig.4 can be factorized.
\par

\begin{figure}[hbt]
\begin{center}
\includegraphics[width=12cm]{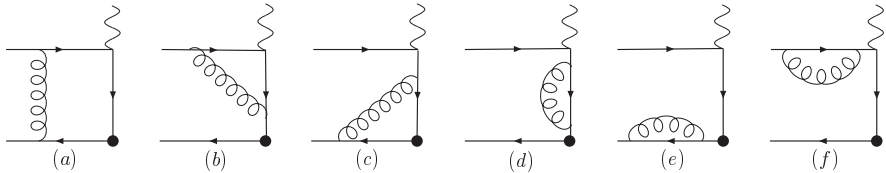}
\end{center}
\caption{Feynman diagrams of the one-loop corrections to Fig.3a.}
\label{Feynman-dg4}
\end{figure}
\par
The contributions except that of Fig.4a can be calculated in a
straightforward way. They are:
\begin{eqnarray}
F(Q^2)\vert_{4e} &=& F(Q^2)\vert_{4f} = F(Q^2)\vert_{3a} \cdot
\frac{\alpha_s}{6\pi}
     \left [ -\ln\frac{\mu^2}{m ^2} -2 \ln\frac{\lambda^2}{m ^2} -4 \right
],
\nonumber\\
F(Q^2)\vert_{4d} &=& F(Q^2)\vert_{3a}\cdot \frac{-\alpha_s}{3\pi} \left [
\ln\frac{\mu^2}{Q^2} -\ln x_0 +1 \right],
\nonumber\\
F(Q^2)\vert_{4b} &=&F(Q^2)\vert_{3a}\cdot \frac{2 \alpha_s}{3\pi} \left [
\frac{1}{2}\ln\frac{\mu^2}{Q^2} + \ln\frac{Q^2}{m ^2} +\frac{1}{2}\ln x_0
\right],
\nonumber\\
F(Q^2)\vert_{4c} &=& F(Q^2)\vert_{3a}\cdot \frac{2 \alpha_s}{3\pi}
\left \{ \frac{1}{\bar x_0} \left [ \frac{1}{2} \ln^2 x_0 +2 {\rm
Li}_2(-\frac{\bar x_0}{x_0})  + \ln\frac{Q^2}{m^2}\ln x_0
\right ]
\right.
\nonumber\\
   && \left. \ \ \  +\frac{1}{2} \ln\frac{\mu^2}{Q^2} - \ln\frac{m^2}{Q^2}
-\frac{2+x_0}{2\bar x_0} \ln x_0  \right \}.
\label{1LF}
\end{eqnarray}
It is obvious that the contributions from Fig.4e and Fig.4f are already
contained in the
contribution of Fig.2a and Fig. 2d of the TMD LCWF, respectively.
\par
The contribution from Fig.4a is difficult to calculate and has a complicated
expression. However, for the purpose of the factorization, we only need
the contribution at the leading power of $Q^2$. We have the result for the
combination:
\begin{eqnarray}
F(Q^2)\vert_{4a} -\frac{1}{Q^2}\phi_+^{(1)}\vert_{1a} \otimes \tilde
S^{(0)}\otimes H^{(0)}\vert_{3a}
  &=& -F(Q^2)\vert_{3a}\frac{2\alpha_s x_0}{3\pi^2 \bar x_0}
     \left [ -\ln x_0 \int d^2 k_\perp \frac{1}{\lambda^2 +k_\perp^2} -\pi \ln x_0
\right.
\nonumber\\
   && \left. +\pi \ln x_0 \left ( \ln \frac{Q^2}{\lambda^2} -1 +\frac{1}{2}
\ln x_0 \right ) \right ]
     \left ( 1 + {\mathcal O} (Q^{-2}) \right),
\label{1LWA}
\end{eqnarray}
In the above expression the I.R. divergent term $\ln\lambda^2$ will
be canceled from the integration of $k_\perp$. The expression
contain an U.V. divergence. This divergence comes from TMD LCWF of
Fig.1a and needs to be subtracted.
\par
Since the contributions from Fig.4e and Fig.4f are already contained
in the corresponding contributions of TMD LCWF, i.e., they are
already subtracted for the hard part, we have only collinear
divergences in $F(Q^2)\vert_{4b}$ and $F(Q^2)\vert_{4c}$, signalled
by $\ln m$, as soft divergences which need to be subtracted.
\par\vskip15pt
\noindent
{\bf 4.2. The Convolution $\phi_+^{(1)} \otimes \tilde S^{(0)} \otimes
H^{(0)}$ }
\par
In this section we present the results for the convolution
$\phi_+^{(1)} \otimes \tilde S^{(0)} \otimes H^{(0)}$. We will see
that the convolution contains the same collinear divergences as
those in $F^{(1)}(Q^2)$. Therefore, with our TMD LCWF the collinear
singularities are correctly factorized. Moreover, the collinear
divergence in a given diagram for $F^{(1)}(Q^2)$ appears exactly in
the same class of diagrams for TMD LCWF. This fact can make it
possible to extend the factorization of collinear divergences beyond
one-loop level. We will also see that there are still some I.R.
divergences in the convolution. Therefore, a soft factor is needed
to subtract them.
\par
We find that it is convenient to present the results in the
following combinations:
\begin{eqnarray}
{\mathcal W}_b &=& \left [ \phi_+\vert_{1b} + \phi_+\vert_{2b}+
\phi_+\vert_{2c} \right ]
  \otimes \tilde S^{(0)} \otimes H^{(0)} \vert_{3a}
\nonumber\\
  &=& \frac{2\alpha_s}{3\pi x_0} \phi_0 \left \{
        \ln\frac{\mu^2}{m^2} +\frac{1}{2}\ln\frac{\zeta^2 x_0^2}{\lambda^2}
        -\frac{1}{4}\ln^2 \frac{\zeta^2 x_0^2}{\lambda^2}
       -\frac{\pi^2}{4} + 1
  + \frac{1}{2} \int \frac{ d^2 q_\perp}{\pi} \frac{1}{q^2_\perp +\lambda^2}
\ln \frac{\zeta^2 x_0^2}{ q^2_\perp +\lambda^2}
     \right \},
\nonumber\\
{\mathcal W}_c &=& \left [ \phi_+\vert_{1c} + \phi_+\vert_{2e}+
\phi_+\vert_{2f} \right ]
  \otimes \tilde S^{(0)} \otimes H^{(0)} \vert_{3a}
\nonumber\\
  &=& \frac{2\alpha_s}{3\pi  x_0}\phi_0  \left \{\ln\frac{\mu^2}{m^2}
+\frac{1}{2}\ln\frac{\zeta^2 \bar x_0^2}{\lambda^2}
        -\frac{1}{4}\ln^2 \frac{\zeta^2 \bar x_0^2}{\lambda^2}
       -\frac{\pi^2}{4} + 1 +\frac{1}{\bar x_0} \left ( 2{\rm
Li}(-\frac{\bar x_0}{x_0}) + \ln  x_0 \ln\frac{\lambda^2}{m^2}\right )
\right.
\nonumber\\
     && \left.
     + \frac{\ln  x_0}{\bar x_0}  \int \frac{ d^2 q_\perp}{\pi}
\frac{1}{q_\perp^2 +\lambda^2}
            + \frac{1}{2} \int\frac{ d^2 q_\perp}{\pi} \frac{1}{q^2_\perp
+\lambda^2} \ln \frac{\zeta^2 \bar x_0^2}{ q^2_\perp +\lambda^2}
     \right \},
\nonumber\\
{\mathcal W}_d &=&  \phi_+\vert_{1d}\otimes \tilde S^{(0)} \otimes H^{(0)}
\vert_{3a}
  = -\frac{2\alpha_s}{3\pi x_0} \int \frac{d^2 q_\perp}{\pi}
\frac{1}{q^2_\perp + \lambda^2}.
\label{1LW}
\end{eqnarray}
Comparing the above with the one-loop correction to the form factor
in Eq.(\ref{1LF}), we see that the collinear singularity in
$F(Q^2)\vert_{4b}$ and $F(Q^2)\vert_{4c}$ is exactly reproduced in
${\mathcal W}_b$ and ${\mathcal W}_c$, respectively. But, in the
convolution $\phi_+^{(1)} \otimes \tilde S^{(0)} \otimes H^{(0)}$
there are also I.R. singularities shown as $\ln \lambda^2$ and I.R.
integrals in Eq.(\ref{1LW}). It should be noted that these integrals
are also U.V. divergent. These U.V. divergences are closely combined
with certain I.R. divergences. Once these I.R. divergences are
subtracted, one can expect that the U.V. divergences are subtracted
too. Because of these divergences, one needs to introduce the soft
factor so that the factorization can be completed, i.e., the hard
part $H^{(1)}$ subtracted from Eq.(\ref{con}) is free from any soft
divergence.
\par\vskip15pt
\noindent
{\bf 4.3. The Soft Factor }
\par
As discussed in the above, the I.R. divergences which need to be
subtracted are from TMD LCWF. In general these divergences can be
analyzed by applying an eikonal approximation. Using the
approximation for those diagrams in Fig.1. one can find that the
I.R. divergences in these diagrams can be reproduced by replacing
the quark lines with eikonal lines. The corresponding diagrams just
represent the contributions to the product of the gauge links:
\begin{eqnarray}
  S_4 (z^-, z_\perp,v_1,v_2) &=&
                \frac{1}{3} {\rm Tr} \langle 0 \vert T \left [
L_{v_2}^\dagger (0,-\infty)
                 L_{u}^\dagger (\infty,0)L_{u} (\infty,z) L_{v_1}
(z,-\infty) \right ] \vert 0\rangle \vert_{z^+ =0},
\nonumber\\
L_{v_{1,2}}(z,-\infty) &=& P \exp \left ( -i g_s \int^{0}_{-\infty} d\lambda
     v_{1,2}\cdot G (\lambda v_{1,2}+ z ) \right ),
\nonumber\\
  S_4 (y, q_\perp,\zeta_1, \zeta_2 ) &=& P^+ \int d z^- d^2 z_\perp e^{i y
P^+ z^- - i \vec q_\perp \cdot \vec z_\perp }
            S_4 (z^-, z_\perp),
\end{eqnarray}
the two vectors $v_{1,2}$ and the parameters $\zeta_{1,2}$ are defined as :
\begin{equation}
v^\mu_{1,2} = (v_{1,2}^+, v_{1,2}^-,0,0), \ \ \ \  v_{1,2}^+ >> v_{1,2}^-,
\ \ \ \ \zeta^2_{1,2} = \frac{2 v^-_{1,2} (P^+)^2}{v^+_{1,2}}.
\end{equation}
these vectors represent the trajectories of $q$ and $\bar q$ in Fig.1,
respectively.
The limit $v_{1,2}^+ >> v_{1,2}^-$ should be understood that we neglect all
terms
with a positive power of $\zeta_{1,2}$. The ordering of the limits is
relevant
for calculating $S_4$. We will take the ordering that the limit $v_{1}^+ >>
v_{1}^-$
is taken before taking the limit $v_{2}^+ >> v_{2}^-$.
\par
We have calculated $S_4$ in perturbation theory. At tree-level we
have
\begin{equation}
  S_4^{(0)} (y, q_\perp,\zeta_1, \zeta_2 ) =\delta(y) \delta^2 (\vec
q_\perp), \ \ \ S_4^{(0)} (z^-, z_\perp,v_1,v_2)=1.
\end{equation}
At one-loop level, the contributions to $S_4$ can be represented by those
diagrams in Fig.1 and Fig.2, where
one replaces the quark line with the gauge link $L_{v_1}$ and the antiquark
line with the gauge link
$L_{v_2}$. The contributions from Fig.1 are:
\begin{eqnarray}
S_4(y,\vec q_\perp,\zeta_1,\zeta_2)\vert_{1a} &=& \frac{2\alpha_s}{3 \pi^2}
\left [
    \frac{ \theta (-y)}{y( q^2_\perp +\lambda^2+\zeta_{1}^2 y ^2)}
     + \frac{\theta (y)} {y (q^2_\perp +\lambda^2 +\zeta^2_2 y^2 )} \right
],
\nonumber\\
S_4(y,\vec q_\perp,\zeta_1,\zeta_2)\vert_{1b} &=& -\frac{2\alpha_s}{3 \pi^2}
\left [ \frac{ \theta (-y) }{q^2_\perp +\lambda^2 +\zeta_1^2 y^2 } \left (
\frac{1}{y} \right )_+
     - \frac{1}{2}\delta (y) \frac{1}{q^2_\perp +\lambda^2}\ln \frac{\zeta^2
  x_0^2 }{q^2_\perp +\lambda^2}
    \right ],
\nonumber\\
S_4(y,\vec q_\perp,\zeta_1,\zeta_2)\vert_{1c} &=& \frac{2\alpha_s}{3 \pi^2}
  \left[ \frac{ 1}{q^2_\perp +\lambda^2+\zeta^2_2 y^2} \theta(y) \left (
\frac{1}{y} \right )_+
      + \frac{1}{2}\delta(y) \frac{ 1}{q^2_\perp +\lambda^2}\ln
\frac{\zeta^2 \bar x_0^2 }{q^2_\perp +\lambda^2} \right ],
\nonumber\\
S_4(y,\vec q_\perp,\zeta_1,\zeta_2)\vert_{1d} &=& -\frac{2\alpha_s}{3 \pi^2}
              \frac{\delta (y) }{q^2_\perp +\lambda^2 }.
\label{S4R}
\end{eqnarray}
The ordering of the limits mentioned above only affects the
contribution $S_4\vert_{1a}$. If we change the ordering,
$S_4\vert_{1a}$ changes its sign. The above expressions, when they
are used to calculate the convolution $\phi_+^{(0)}\otimes \tilde
S^{(1)}\otimes H^{(0)}$, will be taken as distributions for $ -x_0
\leq y \leq 1-x_0$. This can be seen from the convolution
\begin{equation}
\phi_+^{(0)}\otimes \tilde S^{(1)}\otimes H^{(0)}
= \phi_0 \int_0^1 dy dq^2_\perp\tilde S^{(1)}(y,q_\perp)
   \theta(x_0+y) H^{(0)}(x_0+y).
\label{xy}
\end{equation}
In Eq.(\ref{S4R}) we have already take
the limit $\zeta\to\infty$, i.e., $u^+<< u^-$ by taking them as
distributions for $ -x_0 \leq y \leq 1-x_0$.
\par
The contributions from the diagrams corresponding to those in Fig.2 are:
\begin{eqnarray}
S_4(y,\vec q_\perp,\zeta_1,\zeta_2)\vert_{2a} &=& S_4(y,\vec
q_\perp,\zeta_1,\zeta_2)\vert_{2b}
= S_4(y,\vec q_\perp,\zeta_1,\zeta_2)\vert_{2d}
\nonumber\\
&=&S_4(y,\vec q_\perp,\zeta_1,\zeta_2)\vert_{2e}
        = S_4^{(0)}(y,\vec q_\perp,\zeta_1,\zeta_2) \cdot
\frac{\alpha_s}{3\pi}
    \ln \frac{\mu^2}{\lambda^2},
\nonumber\\
S_4(y,\vec q_\perp,\zeta_1,\zeta_2)\vert_{2c} &=& -S^{(0)}_4(y,\vec
q_\perp,\zeta_1,\zeta_2) \frac{\alpha_s}{3\pi}
    \ln \frac{\mu^2}{\lambda^2}\ln\frac{\zeta^2}{\zeta^2_1},
\nonumber\\
S_4(y,\vec q_\perp,\zeta_1,\zeta_2)\vert_{2f} &=& -S^{(0)}_4(y,\vec
q_\perp,\zeta_1,\zeta_2) \frac{\alpha_s}{3\pi}
    \ln \frac{\mu^2}{\lambda^2}\ln\frac{\zeta^2}{\zeta^2_2}.
\end{eqnarray}
$S_4$ has the charge conjugation symmetry $S_4(y,\vec
q_\perp,\zeta_1,\zeta_2) = S_4(-y,-\vec q_\perp,\zeta_2,\zeta_1)$.
We can use the symmetry to simplify our definition of the soft
factor by taking the limit $\zeta_2 \to \zeta_1 =\zeta_v$.
\par
If we identify the soft factor $\tilde S$ with $S_4$ as $\tilde S
(z^-, z_\perp) = [ S_4 (z^-,z_\perp,v_1,v_2)]^{-1}$, the I.R.
divergences in $\phi_+(x,k_\perp, \zeta)$ are indeed subtracted if
we integrate over $x$ and $k_\perp$ and extend the integration
region of $x$ from $-\infty$ to $\infty$. However, in these
convolutions $\phi_+^{(1)}\otimes \tilde S^{(0)}\otimes H^{(0)}$ and
$\phi_+^{(0)}\otimes \tilde S^{(1)}\otimes H^{(0)}$, the integrands
are not $\tilde S^{(1)}$ or $\phi_+^{(1)}$ alone, but multiplied
with $H^{(0)}$, and the integration region of $x$ is restricted from
$0$ to $1$. This results in that the I.R. singularities in
$\phi_+^{(1)}\otimes \tilde S^{(0)}\otimes H^{(0)}$ are not exactly
reproduced in $\phi_+^{(0)}\otimes \tilde S^{(1)}\otimes H^{(0)}$,
and there are end-point singularities in $\phi_+^{(0)}\otimes \tilde
S^{(1)}\otimes H^{(0)}$ when $ y$ approaches to $-x_0$ or $1-x_0$.
The problem is related that the proposed soft factor does not vanish
at the end points, while the TMD LCWF does. Therefore the
convolution with $\tilde S (z^-, z_\perp) = [ S_4
(z^-,z_\perp,v_1,v_2)]^{-1}$ is ill defined and the proposed soft
factor is not right for the case.
\par
To overcome the problem, one can construct quantities from $S_4$ and their
contributions
have a one-to-one correspondence to the one-loop contributions to
$\phi_+(x,k_\perp, \zeta)$
in Eq.(\ref{1LWF}) and Eq.(\ref{1LWFA}).
The correspondence means that the I.R. divergences in Eq. (\ref{1LWF}) and
the U.V. divergences
in Eq.(\ref{1LWFA}) are exactly reproduced. For this we first define:
\begin{eqnarray}
{\mathcal S}_1(y,x, \vec q_\perp,\zeta_v) &=& \left [ 1 - \theta (-y)
\frac{y}{y-x} -\theta(y) \frac{y}{y+1-x} \right ]
\nonumber\\
&& \cdot \frac{1}{2} \lim_{\zeta_2\to\zeta_1=\zeta_v}
\left [S_4(q^+, \vec q_\perp, \zeta_1, \zeta_2)+S_4(q^+,\vec q_\perp,
\zeta_2, \zeta_1) \right ].
\end{eqnarray}
The contributions from Fig.1 to ${\mathcal S}_1$ can be easily obtained from
those of $S_4$. We have:
\begin{eqnarray}
{\mathcal S}_1(y,x, \vec q_\perp,\zeta_v) \vert_{1a} &=&  0,
\nonumber\\
  {\mathcal S}_1(y,x, \vec q_\perp,\zeta_v) \vert_{1b} &=&
   -\frac{2\alpha_s}{3 \pi^2 P^+}
   \left [ \frac{ \theta (x_0 -x) }{q^2_\perp +\lambda^2 +\zeta_v^2
(x-x_0)^2 }
     \frac{x}{x_0}\left ( \frac{1}{x-x_0} \right )_+
\right.
\nonumber\\
   &&\left. \ \ \ \ \ \ \ \ \ \ \ \
    - \frac{1}{2}\delta (x-x_0) \frac{1}{q^2_\perp +\lambda^2}\ln
\frac{\zeta^2  x_0^2 }{q^2_\perp +\lambda^2}
    \right ]
\nonumber\\
{\mathcal S}_1(y,x, \vec q_\perp,\zeta_v)\vert_{1c} &=&
     \frac{2\alpha_s}{3 \pi^2 P^+}
   \left [ \frac{ \theta (x -x_0) }{q^2_\perp +\lambda^2 +\zeta_v^2
(x-x_0)^2 }
     \frac{\bar x}{\bar x_0}\left ( \frac{1}{x-x_0} \right )_+
\right.
\nonumber\\
   &&\left. \ \ \ \ \ \ \ \ \ \ \ \
     + \frac{1}{2}\delta (x-x_0) \frac{1}{q^2_\perp +\lambda^2}\ln
\frac{\zeta^2  \bar x_0^2 }{q^2_\perp +\lambda^2}
    \right ],
\nonumber\\
{\mathcal S}_1(y,x, \vec q_\perp,\zeta_v) \vert_{1d} &=& -\frac{2\alpha_s}{3
\pi^2 P^+} \frac{\delta(x-x_0)}{q^2_\perp +\lambda^2},
\label{1LS1}
\end{eqnarray}
where we have taken $y=x-x_0$ because of Eq.(\ref{xy}). The
contributions from Fig.2 to ${\mathcal S}_1$ are the same as those of $S_4$
by taking
$\zeta_2=\zeta_1=\zeta_v$. Comparing Eq.(\ref{1LS1}) with
Eq.(\ref{1LWF}), we can see that the I.R. singularities in
$\phi_+\vert_{1b}$, $\phi_+\vert_{1c}$ and $\phi_+\vert_{1d}$ are
exactly reproduced ${\mathcal S}_1\vert_{1b}$, ${\mathcal
S}_1\vert_{1c}$ and ${\mathcal S}_1\vert_{1d}$, respectively.
\par
Next we define
\begin{eqnarray}
{\mathcal S}_2(y,x, \vec q_\perp, \zeta_v)  &=&   y
\left [ 1 - \theta (-y) \frac{y}{y-x} -\theta(y) \frac{y}{y+1-x} \right ]
\nonumber\\
&& \cdot \frac{1}{2} \lim_{\zeta_2 \to \zeta_1 =\zeta_v} \left [ S_4(q^+,
q_\perp, \zeta_1,\zeta_2)
  -S_4(q^+, q_\perp, \zeta_2,\zeta_1) \right ].
\end{eqnarray}
One easily finds that ${\mathcal S}_2$ only receives a nonzero contribution
from Fig.1a:
\begin{eqnarray}
{\mathcal S}_2 &=& {\mathcal S}_2\vert_{1a},
\nonumber\\
{\mathcal S}_2 (y,x, \vec q_\perp, \zeta_v)\vert_{1a} &=& \frac{2\alpha_s}{3
\pi^2 P^+}
   \left [ \frac{ \theta (x_0 -x) }{q^2_\perp +\lambda^2 +\zeta_v^2
(x-x_0)^2 }
     \frac{x}{x_0}
     +\frac{ \theta (x -x_0) }{q^2_\perp +\lambda^2 +\zeta_v^2 (x-x_0)^2 }
     \frac{1-x}{1-x_0}
    \right ],
\label{1LS2}
\end{eqnarray}
where we have taken $y=x-x_0$ again.
It is noted that the U.V. divergence in $\phi_+\vert_{1a}$ in
Eq.({\ref{1LWFA})
is reproduced by ${\mathcal S}_2 \vert_{1a}$.
\par

\begin{figure}[hbt]
\begin{center}
\includegraphics[width=7cm]{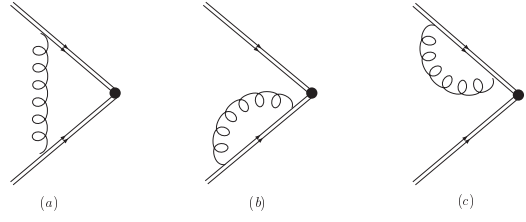}
\end{center}
\caption{One-loop contribution to $S_2$. The double lines represent the two
gauge links. One is for
$L_{v_1}$, the other one is for $L^\dagger_{v_2}$.  }
\label{Feynman-dg5}
\end{figure}
There are extra I.R. divergences introduced by the contributions from Fig.2a
and Fig.2d
for $S_4$. These can be subtracted
by introducing
\begin{equation}
S_2 (\zeta_1,\zeta_2)=\frac{1}{3} {\rm Tr} \langle 0 \vert T \left [
L^\dagger_{v_2}(0,-\infty) L_{v_1}(0,-\infty) \right ] \vert 0 \rangle.
\end{equation}
As for $S_4$, the same limits about the vectors $v_1$ and $v_2$ are taken.
At leading order $S_2^{(0)}=1$. At one-loop level, the contributions are
from the diagrams given in Fig.5.
They are:
\begin{eqnarray}
  S_2\vert_{5a} &=& \frac{\alpha_s}{3\pi}\ln \frac{\mu^2}{\lambda^2} \ln
\frac{\zeta^2_1}{\zeta^2_2},
\nonumber\\
S_2 \vert_{5b} &=& S_2\vert_{5c} = \frac{\alpha_s}{3\pi} \ln
\frac{\mu^2}{\lambda^2}.
\end{eqnarray}
In ${\mathcal S}_1$ we have taken the limit $\zeta_2 \to\zeta_1=\zeta_v$.
Before the limit
there is an I.R. singularity in ${\mathcal S}_{1a}$. This singularity is
exactly the same
as $S_2\vert_{5a}$. We take the limit $\zeta_2 \to\zeta_1=\zeta_v$ for $S_2$
and define:
\begin{equation}
S_{2s} (\zeta_v) =\lim_{\zeta_2 \to \zeta_1 =\zeta_v}
\frac{1}{2} \left ( S_2(\zeta_1, \zeta_2)+S_2( \zeta_2, \zeta_1) \right ).
\end{equation}
The I.R. singularities in Fig.2a and Fig.2d for $S_4$, hence for ${\mathcal
S}_1$ are reproduced
by $S_{2s}$.
It should be noted that the limit $\zeta_2 \to\zeta_1$ for $S_{2s}$ is
nontrivial. If we take $\zeta_1 =\zeta_2$ at the beginning,
we will always have $S_2 =1$.
\par
Finally we can define the correct soft factor as:
\begin{eqnarray}
\tilde S (z^-, z_\perp,x,\zeta_v) &=&
\left (\frac{ {\mathcal S}_1(z^-, z_\perp,x,\zeta_v)+{\mathcal S}_2(z^-,
z_\perp,x,\zeta_v)}{S_{2s}} \right )^{-1},
\nonumber\\
\tilde S (y, x, \vec q_\perp,\zeta_v) &=& P^+\int d z^- d^2 z_\perp e^{i y
P^+ z^- - i \vec q_\perp \cdot \vec z_\perp }
             \tilde S (z^-, z_\perp,x,\zeta_v).
\end{eqnarray}
With this we will show that the extracted hard part is free from any
soft divergence.
\par\vskip15pt
\noindent
{\bf 4.4. The Convolution $\phi_+^{(0)}\otimes \tilde S^{(1)}\otimes
H^{(0)}$ and The Result for $H^{(1)}$}
\par
With the soft factor we define the following combinations of
convolution,
\begin{eqnarray}
{\mathcal U}_b &=&
\phi_+^{(0)}\otimes \left ( \tilde S^{(1)}_{1b} +\tilde S^{(1)}_{2b}+\tilde
S^{(1)}_{2c}\right ) \otimes H^{(0)}_{3a}
\nonumber\\
&=&
     \frac{2\alpha_s}{3\pi x_0} \left \{
       \left [ \frac{\pi^2}{12} + \frac{1}{4} \ln^2
\frac{\lambda^2}{\zeta_v^2 x_0} \right ]
       -\frac{1}{2} \ln\frac{\lambda^2}{\mu^2} \ln \frac{\zeta^2}{\zeta^2_v}
+\frac{1}{2} \ln\frac{\lambda^2}{\mu^2}
  + \frac{1}{2} \int \frac{ d^2 q_\perp}{\pi}\frac{1}{q^2_\perp +\lambda^2}
\ln \frac{ q^2_\perp +\lambda^2}{\zeta^2 x_0^2}
     \right \},
\nonumber\\
{\mathcal U}_c &=&
\phi_+^{(0)}\otimes \left ( \tilde S^{(1)}_{1c} +\tilde S^{(1)}_{2e}+\tilde
S^{(1)}_{2f}\right ) \otimes H^{(0)}_{3a}
\nonumber\\
&=& \frac{2\alpha_s}{3\pi  x_0} \left \{
       \left [ \frac{\pi^2}{12} + \frac{1}{4} \ln^2
\frac{\lambda^2}{\zeta_v^2 \bar x_0^2} \right ]
       -\frac{1}{2} \ln\frac{\lambda^2}{\mu^2} \ln \frac{\zeta^2}{\zeta^2_v}
+\frac{1}{2} \ln\frac{\lambda^2}{\mu^2}
      -\frac{1}{\bar x_0} \left ( 2{\rm Li}(-\frac{\bar x_0}{ x_0}) + \ln
x_0 \ln\frac{\lambda^2}{\zeta_v^2 \bar x_0^2}\right )
\right.
\nonumber\\
    &&\left. - \frac{\ln x_0}{\bar x_0}  \int \frac{ d^2
q_\perp}{\pi}\frac{1}{q_\perp^2 +\lambda^2}
            + \frac{1}{2} \int \frac{ d^2 q_\perp}{\pi} \frac{1}{q^2_\perp
+\lambda^2} \ln \frac{ q^2_\perp +\lambda^2}{\zeta^2 \bar x_0^2}
     \right \},
\nonumber\\
{\mathcal U}_d &=&
\phi_+^{(0)}\otimes  \tilde S^{(1)}_{1d} \otimes H^{(0)}_{3a} =
\frac{2\alpha_s}{3\pi x_0} \int \frac{d^2 q_\perp}{\pi} \frac{1}{q^2_\perp +
\lambda^2}.
\nonumber\\
{\mathcal U}_a &=&
\phi_+^{(0)}\otimes  \tilde S^{(1)}_{1a} \otimes H^{(0)}_{3a}
\nonumber\\
&=& \frac{2\alpha_s}{3\pi} \left \{
-\frac{\ln x_0}{\bar x_0} \int \frac{d^2 q_\perp}{\pi} \frac{1}{q^2_\perp
+\lambda^2}
+2 \ln\frac{\bar x_0}{ x_0} -\frac{1}{\bar x_0} \left [
     2 Li_2(-\frac{\bar x_0}{x_0}) -\ln x_0 \ln\frac{\zeta_v^2 \bar
x_0^2}{\lambda^2} \right ] \right
     \}.
\label{1LU}
\end{eqnarray}
The convolution $\phi_+^{(0)}\otimes \tilde S^{(1)}\otimes H^{(0)}$
is the sum ${\mathcal U}_a+{\mathcal U}_b+ {\mathcal U}_c+ {\mathcal
U}_d$. Comparing Eq.(\ref{1LW}) and Eq.(\ref{1LU}), one can see that
the I.R. divergence and U.V. divergence in ${\mathcal W}_b$,
${\mathcal W}_c$ and ${\mathcal W}_d$ is canceled by ${\mathcal
U}_b$, ${\mathcal U}_c$ and ${\mathcal U}_d$, respectively. The sum
$\sum_{i=b,c,d} \left [{\mathcal U}_i+ {\mathcal W}_i \right ]$
contains only collinear singularities which are the same in
$F^{(1)}(Q^2)$. The U.V. divergence in Eq.(\ref{1LWA}) is also
canceled by that in ${\mathcal U}_a$. Hence, with the soft factor
the hard part $H$ will be infra-red finite. It is clear that the
soft divergences are canceled or subtracted in a diagram-by-diagram
basis, e.g., the collinear singularity in Fig.4b for the form factor
is subtracted by that in Fig.1b and Fig.1c for the TMD LCWF, and the
I.R. singularity in in Fig.1b and Fig.1c for the TMD LCWF is
canceled by that in in Fig.1b and Fig.1c for the soft factor. The
contribution to the TMD LCWF from Fig. 2b  is completely subtracted
by the contribution from Fig.2b to the soft factor. Therefore, after
the subtraction and cancelation the contribution from Fig.4b to the
hard part $H^{(1)}$ is free from any soft divergence.
\par
We now can subtract the contribution from Fig.4 to get $H^{(1)}$:
\begin{eqnarray}
H(x)\vert_{4e} &=& H(x)\vert_{4f} =0,
\nonumber\\
H(x)\vert_{4d} &=& \frac{-\alpha_s}{3\pi x } \left [ \ln\frac{\mu^2}{Q^2}
-\ln x +1 \right],
\nonumber\\
H(x)\vert_{4b} &=& \frac{\alpha_s}{3\pi x} \left \{ -2 +\frac{\pi^2}{3}
+\ln\frac{Q^2}{\zeta^2}
    -\ln x + \frac{1}{2} \ln\frac{\zeta^2}{\zeta_v^2} \left [
\ln\frac{\zeta^2}{\mu^2} +
    \ln\frac{\zeta_v^2}{\mu^2} +4\ln x \right ] \right \},
\nonumber\\
H(x)\vert_{4c} &=& -\frac{\alpha_s}{3\pi x} \left \{ 2 -\frac{\pi^2}{3} +2
\ln (x\bar x) -\frac{1}{\bar x} \ln^2 x
    -\frac{4}{\bar x} {\rm Li}_2 (-\frac{\bar x}{x} )
+\ln\frac{\zeta^2}{Q^2}
\right.
\nonumber\\
&& \left. -\frac{1}{2} \ln\frac{\zeta^2}{\zeta_v^2} \left [
\ln\frac{\zeta^2}{\mu^2} +
    \ln\frac{\zeta_v^2}{\mu^2} +4\ln \bar x \right ]
     +\frac{2}{\bar x} \ln\frac{\zeta_v^2 \bar x^2}{Q^2} \ln x + 3
\frac{x}{\bar x} \ln x \right \},
\nonumber\\
H(x)\vert_{4a} &=& \frac{2\alpha_s}{3\pi \bar x} \left \{ \ln x \left [
\ln\frac{\zeta_v^2 \bar x^2}{Q^2}
   +2 -\frac{1}{2} \ln x  \right ] -2 Li_2 (-\frac{\bar x}{x} )  +2\bar x
\ln\frac{\bar x}{x} \right \}.
\end{eqnarray}
Combining the results from diagrams obtained by reversing the direction of
quark lines
in Fig. 4 we obtain the complete contribution to $H^{(1)}$:
\begin{eqnarray}
H^{(1)}(x,Q^2,\zeta,\zeta_v,\mu^2) &=& \frac{2\alpha_s}{3\pi}
\frac{1}{x\bar x}
\left [ \frac{1}{2} \ln\frac{Q^2}{\mu^2} + \ln\frac{Q^2}{\zeta^2} +
\frac{1}{2}\ln\frac{\zeta^2}{\zeta_v^2}
\left ( \ln\frac{\zeta^2}{\mu^2} +
    \ln\frac{\zeta_v^2}{\mu^2} \right ) +\ln\frac{\zeta^2}{Q^2} \ln (x\bar
x)
\right.
\nonumber\\
    &&\left. -2 \ln\frac{\zeta_v^2}{Q^2} \ln (x\bar x)
     + \ln\frac{\zeta_v^2}{Q^2} \left ( \bar x \ln \bar x + x \ln x\right
)\right]
    + \frac{\alpha_s}{3\pi x\bar x} \left [ -5 -\bar x \ln^2 x
\right.
\nonumber\\
     && \left.   -x \ln^2 \bar x +x \ln x + \bar x \ln \bar x -2 \ln (x\bar
x)
   + 4 \bar x{\rm Li}_2 (x) +4 x {\rm Li}_2 (\bar x)   \right ]     ,
\label{HP}
\end{eqnarray}
The factorization formula with complete parameter-dependence now reads:
\begin{eqnarray}
F(Q^2) &=&  \frac{1}{Q^2 }  \int dx dy  d^2 k_\perp d ^2 l_\perp
\phi_+(x,k_\perp,\zeta,\mu)
          \tilde S (y, x, l_\perp, \zeta,\zeta_v,\mu) \theta (x+y )
\nonumber\\
&&       \cdot  H (x+y, Q^2,\zeta,\zeta_v,\mu)
          +{\mathcal O}(Q^{-4}),
\end{eqnarray}
with
\begin{equation}
H (x, Q^2,\zeta,\zeta_v,\mu)=H^{(0)}(x, Q^2,\zeta,\zeta_v,\mu)+ H^{(1)}(x,
Q^2,\zeta,\zeta_v,\mu)
        = \frac{1}{x\bar x} +H^{(1)}(x, Q^2,\zeta,\zeta_v,\mu),
\end{equation}
from Eq.(\ref{HP}) it is clear that the hard part $H$ does not contain any
soft divergence and
we complete the TMD factorization at one-loop level. Since the soft
divergences are canceled
in a diagram-by diagram basis, the factorization can be argued to hold
beyond one-loop level.
From our results in this section, it is clear that a soft factor is needed
so that the TMD factorization with the TMD LCWF
defined in Eq.(\ref{def}) can be completed.
\par\vskip20pt
\noindent
{\bf 5. Resummation of Large Logarithms}
\par
We note that our factorization formula can be simplified by
introducing $\tilde \phi_+(x, k_\perp,\zeta,\zeta_v, \mu)$:
\begin{eqnarray}
\tilde \phi_+(x, k_\perp,\zeta,\zeta_v, \mu) &=& \int \frac{ d z^- }{2\pi}
  \frac {d^2 z_\perp}{(2\pi )^2}  e^{ik^+z^- - i \vec z_\perp\cdot \vec
k_\perp}
   \tilde S (z^-,z_\perp,\zeta,\zeta_v,\mu)
\nonumber\\
  && \cdot \langle 0 \vert \bar q(0) L_u^\dagger (\infty, 0)
  \gamma^+ \gamma_5 L_u (\infty,z) q(z) \vert \pi^0(P) \rangle\vert_{z^+=0}
.
\label{NF}
\end{eqnarray}
With  $\tilde \phi_+$ the factorization formula reads:
\begin{equation}
F(Q^2) =  \frac{1}{Q^2 }  \int dx  d^2 k_\perp  \tilde
\phi_+(x,k_\perp,\zeta, \zeta_v, \mu)
            H (x, Q^2,\zeta,\zeta_v,\mu)
          +{\mathcal O}(Q^{-4}).
\label{MF}
\end{equation}
In principle one can re-define the TMD LCWF as that given in
Eq.(\ref{NF}). With the redefined TMD LCWF the factorization takes a
simple form. However, we do not intend to do so because it is
unclear if the same re-definition is useful for other processes
whose TMD factorization have been not studied in detail yet. We will
call $\tilde \phi_+$ as a modified TMD LCWF in this work. It should
be kept in mind that the soft factor is there.

With the one-loop result for every factor, we can work out the
evolution equations for the hard part,
\begin{eqnarray}
\mu\frac{\partial}{\partial \mu}\ln H (x,Q^2,\zeta^2,\zeta^2_v,\mu^2) &=&
-\frac{2\alpha_s}{3\pi}
\left [ 1 + 2 \ln\frac{\zeta^2}{\zeta_v^2} \right ]
+{\mathcal O}(\alpha_s^2)
\nonumber\\
\zeta\frac{\partial}{\partial \zeta}\ln H (x,Q^2,\zeta^2,\zeta^2_v,\mu^2)
&=& \frac{2\alpha_s}{3\pi}
\left [ -2 +2 \ln\frac{\zeta^2}{\mu^2} +2 \ln (x\bar x) \right ]+{\mathcal
O}(\alpha_s^2)
\nonumber\\
\zeta_v\frac{\partial}{\partial \zeta_v}\ln H
(x,Q^2,\zeta^2,\zeta^2_v,\mu^2) &=& \frac{2\alpha_s}{3\pi}
\left [ - 2\ln\frac{\zeta_v^2}{\mu^2}
-4 \ln (x\bar x) +2\left ( \bar x \ln \bar x + x \ln x\right )  \right ]
+{\mathcal O}(\alpha_s^2).
\end{eqnarray}
The corresponding evolutions of $\tilde \phi_+$ reads:
\begin{eqnarray}
\mu\frac{\partial}{\partial \mu}\ln \tilde\phi_+ (x,k_\perp,
\zeta^2,\zeta^2_v,\mu^2) &=& \frac{2\alpha_s}{3\pi}
\left [ 1 + 2 \ln\frac{\zeta^2}{\zeta_v^2} \right ]
+{\mathcal O}(\alpha_s^2)
\nonumber\\
\zeta\frac{\partial}{\partial \zeta} \ln \tilde\phi_+ (x,k_\perp,
\zeta^2,\zeta^2_v,\mu^2) &=& \frac{2\alpha_s}{3\pi}
\left [ 2 - 2 \ln\frac{\zeta^2}{\mu^2} -2 \ln(x\bar x)  \right ]+{\mathcal
O}(\alpha_s^2),
\end{eqnarray}
the evolution with $\zeta_v$ does not take a simple form, it takes a
convolution form:
\begin{equation}
\zeta_v\frac{\partial}{\partial \zeta_v} \tilde\phi_+ (x,k_\perp,
\zeta^2,\zeta^2_v,\mu^2)
=\int_0^1 d y C_v (x,y,\zeta_v, \mu) \tilde\phi_+ (y,k_\perp,
\zeta^2,\zeta^2_v,\mu^2),
\end{equation}
with
\begin{eqnarray}
C_v (x,y) &=& \frac{2\alpha_s}{3\pi} \left \{ \delta(x-y) \left (
2\ln\frac{\zeta_v^2 y\bar y}{\mu^2} -4 \right)
   -2 \left [ \frac{1}{x-y} \left ( \theta(y-x) \frac{x}{y} -\theta(x-y)
\frac{\bar x}{\bar y} \right ) \right ]_+
\right.
\nonumber\\
   && \left. + 2 \left ( \theta(y-x) \frac{x}{y} +\theta(x-y) \frac{\bar
x}{\bar y} \right ) \right \},
\end{eqnarray}
where the last line comes from the box diagram. The $+$-prescription is for
$x$.
It should be noted that:
\begin{eqnarray}
\int_0^1 dx \frac{1}{x\bar x}
\left [ \frac{1}{x-y} \left ( \theta(y-x) \frac{x}{y} -\theta(x-y)
\frac{\bar x}{\bar y} \right ) \right ]_+
& =&  -\frac{1}{y\bar y}\left [ 2+  \ln (y\bar y)\right ].
\nonumber\\
\int_0^1 dx \frac{1}{x\bar x}\left ( \theta(y-x) \frac{x}{y} +\theta(x-y)
\frac{\bar x}{\bar y} \right )
&=& -\frac{1}{y\bar y} \left ( \bar y \ln \bar y + y\ln y \right ).
\end{eqnarray}
With the above equations one can easily check that the form factor does not
depend on $\mu,\zeta$ and $\zeta_v$,
as expected.
\par
In Eq.(\ref{HP}) there are different possibly large logarithms.
Beside those resulted from different scales like $\zeta^2$,
$\zeta_v^2$... etc., there are large logarithms around the end
points of $x$, like $\ln^2 x$ and $\ln x$ if $x$ goes to $0$, and
$\ln^2 (1-x)$ and $\ln (1-x)$ if $x$ goes to $1$. These logarithms
also appear in the collinear factorization. They make the
perturbative expansion of $H$ unreliable and must be resummed. With
the TMD factorization, the resummations can be accomplished in a
simple way.
\par
The result for $H$ can be simplified by noting that the modified TMD
LCWF is symmetric under the exchange $x \leftrightarrow 1-x$. It is
convenient to introduce a parameter $\rho$ to relate $\zeta_v^2$ and
$\zeta^2$:
\begin{equation}
\zeta_v^2 = \frac{u^2 v^2}{4 (v\cdot u)^2} \zeta^2=\zeta^2/\rho^2, \ \ \ \ \
\rho =\sqrt{\frac{v^+ u^-}{v^- u^+}}.
\end{equation}
The factorization in Eq.(\ref{MF}) now reads:
\begin{equation}
F(Q^2) =\frac{1}{Q^2 }  \int dx d^2 k_\perp \tilde \phi_+ (x, k_\perp,
\zeta,\rho,\mu) H (x, Q, \zeta,\rho,\mu)
          +{\mathcal O}(Q^{-4}).
\end{equation}
and
\begin{eqnarray}
H^{(1)}(x,Q,\zeta,\rho,\mu) &=& \frac{2\alpha_s}{3\pi x\bar x}
\left [ \frac{1}{2} \ln\frac{Q^2}{\mu^2} + \ln\frac{Q^2}{\zeta^2} -
\frac{1}{2} \ln^2 \rho^2
+\ln\rho^2 \ln\frac{\zeta^2}{\mu^2}
\right.
\nonumber\\
    &&\left. -2 \ln\frac{\zeta^2}{Q^2}\left (\ln x - x \ln x \right )
     +2 \ln \rho^2 \left ( 2\ln x - x \ln x \right ) \right]
\nonumber\\
   && + \frac{2\alpha_s}{3\pi x\bar x} \left [ -\frac{5}{2}  - \bar x \ln^2
x
         +x \ln x  - 2 \ln x
+ 4 \bar x{\rm Li}_2 (x)   \right ],
\label{HH}
\end{eqnarray}
where we have used the property
$\tilde \phi_+ (x, k_\perp, \zeta,\rho,\mu) =\tilde \phi_+ (1-x, k_\perp,
\zeta,\rho,\mu)$
to simplify the hard part $H$. The evolution of $H$ with various scales
reads:
\begin{eqnarray}
\mu\frac{\partial}{\partial \mu}\ln H (x,Q,\zeta,\rho,\mu) &=&
-\frac{2\alpha_s}{3\pi}
\left [ 1 +2 \ln\rho^2 \right ]
+{\mathcal O}(\alpha_s^2)
\nonumber\\
\zeta\frac{\partial}{\partial \zeta}\ln H (x,Q,\zeta,\rho,\mu) &=&
-\frac{4\alpha_s}{3\pi}
\left [ 1 -\ln\rho^2  +2 \ln x -2  x \ln x \right ]+{\mathcal O}(\alpha_s^2)
\nonumber\\
\rho\frac{\partial}{\partial \rho}\ln H (x,Q,\zeta,\rho,\mu) &=&
\frac{4\alpha_s}{3\pi}
\left [  -\ln\rho^2 + \ln\frac{\zeta^2}{\mu^2}
+ 4 \ln x  -2 x \ln x  \right ]
+{\mathcal O}(\alpha_s^2).
\label{RGE1}
\end{eqnarray}
\par
To avoid large logarithms in the wave function, we first choose
appropriate scales to eliminate them
\begin{equation}
F(Q^2) =\frac{1}{Q^2 } \lim_{b\to 0}  \int_0^1 dx \tilde \phi_+
(x,\zeta_L,\rho_L,\mu_L) H (x, Q, \zeta_L,\rho_L,\mu_L)
          +{\mathcal O}(Q^{-4}).
\end{equation}
If we use the result in Eq.(\ref{HH}), there will be large
logarithms in $H (x, Q, \zeta_L,\rho_L,\mu_L)$, especially terms
like $\ln^2 x$, $\ln x$ when $x\to 0$. It is possible to resum those
large logarithms terms by using the evolutions in Eq.(\ref{RGE1}) to
express $H (x, Q, \zeta_L,\rho_L,\mu_L)$ as:
\begin{equation}
H (x, Q, \zeta_L,\rho_L,\mu_L) = e^{-S}H (x, Q, \zeta_H,\rho_H,\mu_H).
\end{equation}
This represents the resummed form for the hard part.
By taking appreciated scales $\zeta_H$, $\rho_H$ and $\mu_H$ so that
$ H (x, Q, \zeta_H,\rho_H,\mu_H)$ does not contain any large logarithms. It
should be noted
that $\ln \rho_L$ is not a large logarithms. We first use the
$\mu$-evolution to evolute
the hard part to a high scale $\mu_H$ and take $\mu_H =Q$.
As the next we use the
$\zeta$-evolution to evolute $H$ to a high scale $\zeta_H =Q$. After these
the hard part reads:
\begin{eqnarray}
H (x, Q, \zeta_L,\rho_L,\mu_L) &=& \exp \left \{ - \int_{\mu_H}^{\mu_L}
\frac{d\mu}{\mu}
   \cdot \frac{2\alpha_s(\mu)}{3\pi} \left ( 1 + 2\ln\rho_L^2 \right )
\right \}
\nonumber\\
    && \cdot  \exp \left \{ -\frac{2 \alpha_s(Q)}{3\pi}
   \left [ 1 -\ln\rho_L^2  +2\bar x \ln x  \right ]
\ln\frac{\zeta_H^2}{\zeta_L^2} \right \}
   \cdot
   H (x, Q, \zeta_H,\rho_L,Q),
\nonumber\\
H(x,Q, \zeta_H=Q,\rho_L,Q ) &=& \frac{1}{x\bar x} \left \{ 1 +
\frac{2\alpha_s(Q) }{3\pi }
\left [ -\frac{1}{2} \ln^2 \rho_L^2
     +2 \ln \rho_L^2 \left ( 2\ln x - x \ln x \right )
\right. \right.
\nonumber\\
   && \left. \left.  -\frac{5}{2}  - \bar x \ln^2 x
         +x \ln x  - 2 \ln x
+ 4 \bar x{\rm Li}_2 (x)   \right ] \right \}.
\end{eqnarray}
Now the hard part only contains possibly large logarithms $\ln^2 x$ and $\ln
x$ when $x$ goes to zero.
To resum these logarithms we note that
\begin{eqnarray}
&& -\frac{1}{2} \ln^2 \rho^2 +4 \ln\rho^2 \ln x -\ln^2 x -2 \ln x =
-\frac{1}{2}\left [
\ln^2 \frac{\rho^2}{x^4} -\ln^2\frac {\rho_H^2}{x^4} \right ] -\frac{1}{7},
\nonumber\\
&& \rho_H^2 =\rho^2_H(x) = x^{4+\sqrt{14}} e^{-\sqrt{2/7} }.
\end{eqnarray}
Using the $\rho$-evolution we have:
\begin{eqnarray}
H (x, Q, \zeta_L,\rho_L,\mu_L) &=& \exp (-S) \cdot H (x, Q, Q,\rho_H,Q),
\nonumber\\
S &=&  \frac{2 \alpha_s(Q)}{3\pi}\left \{
   \left [ 1 - \ln\rho_L^2  +2\bar x \ln x  \right ]
\ln\frac{\zeta_H^2}{\zeta_L^2}
   + \ln^2 x +2 \ln x -\frac{1}{7}-2x \ln x\ln \rho_H^2
\right.
\nonumber\\
  &&\left.  +\frac{1}{2} \ln^2 \rho_L^2 - \ln\rho_L^2 ( 4 \ln x -2 x \ln x )
   \right \} -\int^{Q}_{\mu_L} \frac{d\mu}{\mu}
   \cdot \frac{2\alpha_s(\mu)}{3\pi} \left ( 1 + 2\ln\rho_L^2 \right )
\nonumber\\
H(x,Q, Q,\rho_H,Q ) &=& \frac{1}{x\bar x} \left \{ 1 +  \frac{2\alpha_s(Q)
}{3\pi }
\left [ -\frac{1}{7}
     -2 x \ln \rho_H^2 \ln x + x \ln^2 x
\right. \right .
\nonumber\\
&& \left.\left.
\ \ \ \ \  +x \ln x + 4 \bar x{\rm Li}_2 (x)-\frac{5}{2}   \right ] \right
\}.
\end{eqnarray}
Now there is no large logarithms in $H(x,Q, Q,\rho_H,Q )$. It is
instructive to compare the result of $H (x, Q,
\zeta_L,\rho_L,\mu_L)$ with and without the resummation for $x\to
0$. Without the resummation the result for $x\to 0$ reads:
\begin{equation}
x H (x, Q, \zeta_L,\rho_L,\mu_L) \sim  1 -\frac{2\alpha_s(\mu_L)}{3\pi}
\ln^2 x,
\end{equation}
therefore the one-loop correction can be very large and is divergent. After
the resummation
the result for $x\to 0$ reads:
\begin{equation}
x H (x, Q, \zeta_L,\rho_L,\mu_L) \sim  x^{-\frac{2\alpha_s(Q)}{3\pi} \ln x},
\end{equation}
which will approaches to $0$ faster than any positive power of $x$.
\par\vskip20pt
\noindent
{\bf 6. Summary}
\par
In this work we start with a consistent definition of TMD LCWF's to
explicitly show that the TMD factorization for the process $\gamma^*
\pi^0 \to \gamma$ can be made at one-loop level, where a soft factor
must be implemented. The soft factor is defined as a product of
gauge links. It makes the hard part free from I.R. singularities. In
contrast to the corresponding collinear factorization, where the
amplitude is factorized as a convolution with the standard
light-cone wave function and a hard part, the amplitude in the TMD
factorization is factorized as a convolution of a TMD light-cone
wave function, a soft factor and a hard part. From our results, the
cancelation of all soft divergences happens in a diagram-by-diagram
way, hence it is possible to extend the factorization beyond
one-loop level.
\par
With the factorization, the hard part can be safely calculated in a
perturbative expansion in $\alpha_s$. But in general results from
perturbative expansions have large logarithms, which can make the
results unreliable. A resummation of these large logarithms is
needed. With the TMD factorization we show that the resummation in
our case can be done simply. It should be mentioned that it is
possible to show that the two factorization are equivalent, as the
study of TMD factorization for the radiative leptonic decay of
$B$-meson shows\cite{MW2}.
\par
After the detailed examination of TMD factorization for the simple
cases involving one hadron only, we can proceed to study more
complicated cases involving more hadrons, particularly $B$-decays
into light hadrons.
\par
\vskip 5mm
\par\noindent
{\bf\large Acknowledgments}
\par
The authors would like to thank Prof. X.D. Ji for discussions. This
work is supported by National Nature Science Foundation of P.R.
China(No. 10421003 and No. 10575126).
\par


\end{document}